\documentclass[final]{elsarticle}

\usepackage{amssymb,amsmath,amsthm}
\usepackage{subfigure}
\usepackage{graphicx,url,color}
\graphicspath{{./Figures/}}

\usepackage{url}
\urldef{\mailsa}\path|{alexgc, sandu, sturler, emconsta}@vt.edu|

\newcommand{\Jfunc}{\mathcal J }
\newcommand{\Model}{\mathcal M}
\newcommand{\M}{\mathbf{M}}

\newcommand{\Hobs}{\mathcal{H}}
\newcommand{\HH}{\mathbf H}

\newcommand{\B}{\mathbf{B}}
\newcommand{\C}{\mathbf{C}}
\newcommand{\R}{\mathbf{R}}

\newcommand{\x}{   \mathbf{x} }
\newcommand{\xb}{ \mathbf{x}^{\rm b} }
\newcommand{\xa}{ \mathbf{x}^{\rm a} }

\newcommand{\xv}{ \mathbf{x}^{\rm v} }
\newcommand{\y}{ \mathbf{y} }

\renewcommand{\u}{ \mathbf{u} }

\begin{document}

\thispagestyle{empty}
\setcounter{page}{0}

\begin{Huge}
\begin{center}
Computational Science Laboratory Technical Report CSL-TR-2-2013 \\
\today
\end{center}
\end{Huge}
\vfil
\begin{huge}
\begin{center}
Alexandru Cioaca, \\ Adrian Sandu, and Eric de Sturler
\end{center}
\end{huge}

\vfil
\begin{huge}
\begin{it}
\begin{center}
Efficient methods for computing observation impact \\
in 4D-Var data assimilation
\end{center}
\end{it}
\end{huge}
\vfil

\begin{large}
\begin{center}
Computational Science Laboratory \\
Computer Science Department \\
Virginia Polytechnic Institute and State University \\
Blacksburg, VA 24060 \\
Phone: (540)-231-2193 \\
Fax: (540)-231-6075 \\ 
Email: \url{sandu@cs.vt.edu} \\
Web: \url{http://csl.cs.vt.edu}
\end{center}
\end{large}

\vspace*{1cm}

\begin{tabular}{ccc}
\includegraphics[width=2.5in]{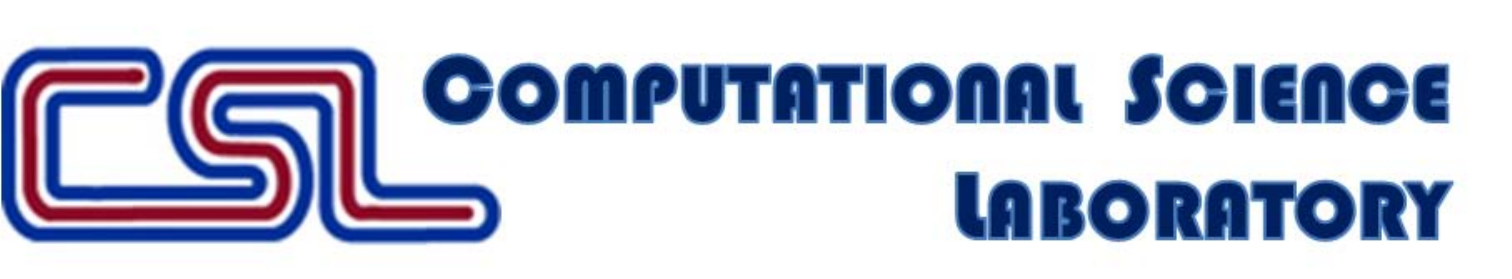}
&\hspace{2.5in}&
\includegraphics[width=2.5in]{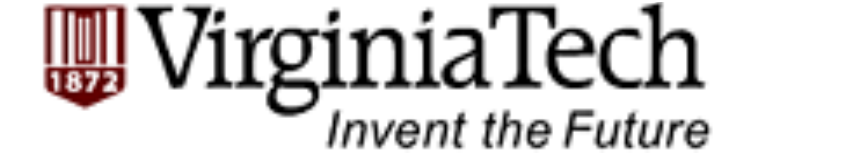} \\
{\bf\em Innovative Computational Solutions} &&\\
\end{tabular}

\newpage

\title{Efficient methods for computing observation impact \\
in 4D-Var data assimilation}

\author{Alexandru Cioaca}
\author{Adrian Sandu}
\author{Eric de Sturler}


\begin{abstract}

This paper presents a practical computational approach to quantify the effect of individual observations in estimating the state of a system. Such an analysis can be used for pruning redundant measurements, and for designing future sensor networks.

The mathematical approach is based on computing the sensitivity of the reanalysis (unconstrained optimization solution) with respect to the data. The computational cost is dominated by the solution of a linear system, whose matrix is the Hessian of the cost function, and is only available in operator form. The right hand side is the gradient of a scalar cost function that quantifies the forecast error of the numerical model. The use of adjoint models to obtain the necessary first and second order derivatives is discussed. We study various strategies to accelerate the computation, including matrix-free iterative solvers, preconditioners, and an in-house multigrid solver. Experiments are conducted on both a small-size shallow-water equations model, and on a large-scale numerical weather prediction model, in order to illustrate the capabilities of the new methodology. 

\end{abstract}


\maketitle

\section{Introduction}

Data assimilation is the process that combines prior information, numerical model predictions, observational data, and the corresponding error statistics, to produce a better estimate of the state of a physical system.  In this paper we consider the four dimensional variational (4D-Var) approach, which formulates data assimilation as a nonlinear optimization problem constrained by the numerical model. The initial conditions (as well as boundary conditions, forcings, or model parameters) are adjusted such as to minimize the discrepancy between the model trajectory and a set of time-distributed observations. In real-time operations, the analysis is performed in cycles: observations within an assimilation time window are used to obtain an optimal trajectory, which provides the initial condition for the next time window, and the process is repeated.

The quality and availability of observational data have a considerable impact on the accuracy of the resulting reanalysis (optimal initial conditions).
We are interested to quantify rigorously the impact that different observations have on the result of data assimilation. The assessment of contributions of observations  has important applications such as detecting erroneous data (e.g., due to faulty sensors), pruning redundant or unimportant data, and finding the most important locations where future sensors should be deployed.

Early studies of observation impact were concerned with quantifying the predictability of the numerical model, using breeding vectors, potential vorticity and singular vectors \cite{Palmer_1999, Sandu_HSV}. It was assumed that observations in areas of high uncertainty would significantly improve the reanalysis, which led to the concept of targeted and adaptive observations. Later research developed specialized methods such as ensemble transformation techniques \cite{Bishop_1999, Bishop_2001} and adjoint-based model sensitivity \cite{Bergot_2001, Fourrie_2002}. Some of this research was validated through Observing System Simulation Experiments (OSSEs) \cite{FASTEX, NORPEX, THORPEX}. Recent research shifted focus from the numerical model to studying the entire data assimilation system for ensemble-based methods \cite{Kalnay_2008}, 3D-Var \cite{Daley_2000}, nonlinear 4D-Var \cite{Langland_2004,NavonDaescu} and incremental 4D-Var \cite{Tremolet_INCDA}.  Important alternative approaches to asses the importance of 
observations are based on statistical design \cite{Berliner_1999} and information theory \cite{ZupInfoTheory,Singh_2012}. 

The focus of this work is on the sensitivity of the 4D-Var reanalysis to observations. The sensitivity equations are derived rigorously in the theoretical framework of optimal control and optimization \cite{LeDimet_1997,Daescu_2008,Daescu_2010}. Sensitivity analysis reveals subsets of data, and areas in the computational domain, which have a large contribution in reducing (or increasing) the forecast error.  The solution of the 4D-Var sensitivity equations involves the solution of a linear system, whose system matrix is the Hessian of the 4D-Var cost function. This matrix is typically very large and available only in the form of matrix-vector products. 


This work addresses two challenges associated with computing sensitivities to observations. The first challenge is the computation of the required first and second order derivatives. The solution discussed herein is based on first and second order adjoint models. The second challenge is obtaining an accurate solution of the large linear system that defines the sensitivities. 
Computational time is an important consideration, especially in applications where the solution is needed real-time. Several solutions are proposed in this work.  A set of preconditioners is selected and tested to speed up the convergence of Krylov solvers. A multigrid strategy is also considered. Tests are conducted using two numerical models. The first one is the 2D shallow water equations, for which all the derivatives can be computed very accurately. The second test is the Weather Research and Forecast (WRF) model, widely used in numerical weather prediction. The experimental results illustrate the potential of the proposed computational approaches to speed up observation impact calculations in real life applications.

The paper is organized as follows. Section \ref{sec:da} reviews the 4D-Var data assimilation approach. Section \ref{sec:obs} covers the theoretical framework of sensitivity analysis in the context of 4D-Var, and derives the equations for the sensitivities to observations. Section \ref{sec:swe} discusses practical computational algorithms and their application to the shallow water equations. Section \ref{sec:wrf} presents the results obtained with the large-scale Weather Research and Forecast (WRF) model. A qualitative discussion of the results is provided in Section \ref{sec:vis}. Conclusions are drawn in Section \ref{sec:end}, and several directions of future research are highlighted.

\section{Data Assimilation}\label{sec:da}

Data assimilation (DA) is the process by which measurements are used to constrain model predictions \cite{Daley_1991,Kalnay_2002}. For this, three sources of information are combined: an a priori estimate of the state of the system (the ``background''), knowledge of the physical laws governing the evolution of the system (captured by the numerical model), and sparse observations of the system. In four dimensional variational (4D-Var) assimilation an optimal initial state $\xa_0$ (the ``reanalysis'') is obtained by minimizing the cost function
\begin{subequations}
\begin{eqnarray}
 \label{eqn:cost_fcn_assim}
 \Jfunc(\x_0) &=& \frac{1}{2} \left( \x_0 - \xb_0 \right)^T \cdot \B_0^{-1} \cdot ( \x_0 - \xb_0 )   \\
            & & \, + \frac{1}{2} \sum_{k=0}^{N} \left( \Hobs_k (\x_k) - \y_k \right)^T \cdot \R_k^{-1} \cdot \left( \Hobs_k (\x_k) - \y_k \right)\,, 
 \nonumber \\
 \xa_0 &=& \arg\min_{\x_0} \Jfunc(\x_0) \,.
 \label{eqn:da}
\end{eqnarray}
\end{subequations}
The first term of the sum \eqref{eqn:cost_fcn_assim} quantifies the departure of the solution from the background state $\xb_0$ at the initial time $t_0$. The term is scaled by the inverse of the background error covariance matrix $\B_0$. The second term measures the mismatch between the forecast trajectory and the observations $\y_k$, which are taken at times $t_0,\dots,t_N$ inside the assimilation window. When assimilating observations only at the initial time $t_0$, the method is known as three dimensional variational (3D-Var), as the additional ``time'' dimension is not present. $\Model$ is the numerical model used to evolve the state vector $\x$ in time. $\Hobs_k$ is the observation operator at assimilation time $t_k$, and maps the discrete model state $\x_k \approx \x(t_k) = \Model_{t_0 \rightarrow t_k} (\x_0)$ to the observation space. $\R_k$ is the observations error covariance matrix. The weighting matrices $\B_0$ and $\R_k$ need to be predefined in order to have a fully-defined problem, and their 
quality 
influences the accuracy of the resulting reanalysis. 

Since an analytical solution for the equation \eqref{eqn:da} is not possible, the minimizer is computed iteratively using numerical optimization methods. Such methods typically require the gradient of the cost function, while Newton-type methods also require second-order derivative information. Higher-order information can be computed using techniques from the theory of adjoint sensitivity analysis \cite{Cacuci_1981}. In this case, first-order adjoint models provide the gradient of the cost function, while second-order adjoint models provide the Hessian-vector product. The methodology of building and using various adjoint models for optimization, sensitivity analysis, and uncertainty quantification can be found in \cite{SanduADJ_2005,Cioaca_2011}.

When 4D-Var is employed in an operational setting (in real time), the reanalysis \eqref{eqn:da} has to be determined within a given time limit, and the iterative solver is stopped after a certain number of iterations, typically before complete convergence. Although the most significant decrease in the cost function usually happens during the first iterations, it is likely the analysis is approximate and does not satisfy exactly the optimality conditions. Slow convergence is a known issue for the solution of highly nonlinear problems of PDE-constrained optimization. The resulting reanalysis can be interpreted as only partially assimilating the observations. Along with the problem of correctly defining the error statistics, it represents one of the practical challenges of data assimilation.

\section{Sensitivity of the Analysis to Observations}\label{sec:obs}

The sensitivity of the analysis to observations is derived in the context of unconstrained optimization, and the presentation follows \cite{Daescu_2008}. Consider the problem of finding a vector $\x = (x_1, x_2, ..., x_n)^T \in \mathbb{R}^n$ that minimizes the twice continuously differentiable cost function
\[
 \min_{\x} \Jfunc(\x,\u) \,.
\]
The function also depends on the vector of parameters $\u \in \mathbb{R}^m$. The implicit function theorem applied to the first order optimality condition 
\begin{equation}
\nabla_\x \, \Jfunc(\bar{\x},\bar{\u}) = 0 
\label{eqn:fooc}
\end{equation}
guarantees there exists a vicinity of $\bar{\u}$ where the optimal solution is a smooth function of the input data, $\x = \x(\u)$ and
$\nabla_\x \, \Jfunc(\x(\u),\u) = 0$. The sensitivity of the optimal solution with respect to the parameters
\[
\nabla_\u \, \x = (\nabla_\u \x_1, \nabla_\u \x_2, ..., \nabla_\u \x_n)  \in \mathbb{R}^{m \times n}
\]
can be expressed as
\begin{equation}
\nabla_\u\, \x(\u) = -\nabla_{\u,\x}^2 \Jfunc (\u, \x) \cdot \left[\nabla_{\x_0,\x_0}^2 \Jfunc(\u, \x)\right] ^{-1}\,.
\label{eqn:diff}
\end{equation}
Consider now a scalar functional $\mathcal{E}$ that represents some quantity of interest of the optimal solution,  $\mathcal{E}(\x(\u))$.
Using chain rule differentiation we obtain its sensitivity to parameters 
\begin{equation}
\nabla_\u \mathcal{E} = \nabla_\u \x \cdot \nabla_\x \mathcal{E} = -\nabla_{\u,\x}^2 \Jfunc \cdot (\nabla_{\x_0,\x_0}^2 \Jfunc)^{-1} \cdot \nabla_\x \mathcal{E}\,.
\end{equation}
For the 4D-Var cost function \eqref{eqn:cost_fcn_assim} the first-order necessary condition reads
\begin{equation}
\nabla_{\x_0}\, \Jfunc(\xa_0) = \B_0^{-1} \left(\xa - \xb\right) + \sum_{k=1}^{N}  \M_{0,k}^T \HH_k^T \R_k^{-1} \left( \Hobs_k(\x_k) - \y_k \right) = 0 \,,
\label{eqn:4dvarfooc}
\end{equation}
where $\M_{0,k} = (\Model_{t_0 \rightarrow t_k})'$ is the tangent linear propagator associated with the numerical model $\Model$, and $\HH_k=(\Hobs_k)'$ is the tangent linear approximation of the observation operator. Differentiating (\ref{eqn:4dvarfooc}) with respect to observations $\y_k$ yields
\begin{equation}
\nabla_{\y_k, \x_0}^2 \, \Jfunc(\xa_0) = -\R_k \, \HH_k\, \M_{0,k}\,,
\end{equation}
which then provides the following analysis sensitivity to observations
\begin{equation}
\nabla_{\y_k}\, \xa_0 = \R_k^{-1}\, \HH_k\, \M_{0,k}\, \left(\nabla_{\x_0,\x_0} \Jfunc(\xa_0)\right) ^{-1} \,.
\end{equation}
In the context of data assimilation we consider $\mathcal{E}(\xa)$ to be a forecast score, i.e., a performance metric for the quality of the reanalysis. If the 4D-Var problem is defined and solved correctly, and if the data is accurate, then the reanalysis $\xa$ should provide a better forecast than the background $\xb$; this is quantified by $\mathcal{E}(\xa) \le \mathcal{E}(\xb)$. Validating the forecast against a reference solution is often used as a way to assess the quality of the initial condition. Since one does not have access to the state of the real system, the reanalysis is verified against another solution of higher accuracy (the ``verification'' forecast). Specifically, we define the forecast score as
\begin{equation}
\label{eqn:forecast-score}
 \mathcal{E}(\xa_0) = (\xa_\textrm{f}-\xv_\textrm{f})^T \, \C\,  (\xa_\textrm{f} - \xv_\textrm{f})
\end{equation}
where $\xa_\textrm{f} = \Model_{t_0\rightarrow t_\textrm{f}}(\xa_0)$ is the model forecast at verification time $t_\textrm{f}$, $\xv_\textrm{f}$ is the verification forecast at $t_\textrm{f}$, and $C$ is a weighting matrix that defines the metric in the state space. For example, $C$ could restrict $\mathcal{E}$ to a subset of grid points, in which case we will quantify the influence of assimilated observations in reducing the forecast error in the corresponding subdomain.

Using the chain rule differentiation for the forecast score we obtain
\[
\nabla_{\y_k} \mathcal{E}(\xa_0) = \nabla_{\y_k} \xa_0 \cdot \nabla_{\xa_0} \mathcal{E}(\xa_0)\,.
\]
This leads to the following expression for the forecast sensitivity to observations
\begin{equation}
\nabla_{\y_k} \mathcal{E}(\xa_0) = \R_k^{-1}\, \HH_k\, \M_{0,k}\, \left(\nabla_{\x_0,\x_0} \Jfunc(\xa_0)\right)^{-1}\, \nabla_{\xa_0} \mathcal{E}(\xa_0)\,.  
\label{eqn:sensobs}
\end{equation}

Obtaining the sensitivity \eqref{eqn:sensobs} is the main goal of this paper. We summarize the big picture from a systems theory perspective. Data assimilation takes as inputs the following parameters: the background estimate of the state of the atmosphere, the observations, the error statistics, and the forecast model. It produces a better initial condition. We perform a forecast using this new estimate, and compute a metric of the forecast error as the mismatch against a verification forecast. We trace back the reduction of the forecast error to the input parameters (specifically, to the observations). This process involves the following three computational steps. 

\subsection{Forecast sensitivity to reanalyzed initial condition}\label{sub:rhs}

We first compute the sensitivity of the forecast score \eqref{eqn:forecast-score} to the optimal initial condition:
\begin{equation}
\label{eqn:score-to-ini}
 \nabla_{\xa_0} \mathcal{E}(\xa_0) = \M_{0,\textrm{f}}^T  \cdot \nabla_{\xa_\textrm{f}} \mathcal{E}(\xa_0)= 2\, \M_{0,\textrm{f}}^T\cdot \C\cdot  (\xa_\textrm{f} - \xv_\textrm{f})\,.
\end{equation}
The gradient \eqref{eqn:score-to-ini} is computed by running the first-order adjoint model, initialized with the forecast error $\xa_\textrm{f} - \xv_\textrm{f}$. The first-order adjoint model evolves the forecast error field backward in time to produce a field of sensitivities at the initial time. This calculation reveals regions in the initial condition to which the output (forecast error, in this case) is most sensitive. This step requires just one adjoint model run and does not add a significant computational load to the method as a whole. 

\subsection{Forecast sensitivity through the 4D-Var system}

%
The second step consists in solving a large-scale linear system of the form:
\begin{equation}
 \nabla_{\x_0,\x_0}^2 \mathcal \Jfunc(\xa_0) \cdot \mu_0 = \nabla_{\xa_0} \mathcal{E}(\xa_0)\,.
 \label{eqn:linsys}
\end{equation}
The system matrix is the Hessian of the 4D-Var cost function evaluated at the reanalysis. The right-hand side is the vector of sensitivities \eqref{eqn:score-to-ini}. The linear system \eqref{eqn:linsys} solves the matrix-vector product $(\nabla_{\x_0,\x_0}^2 \mathcal \Jfunc)^{-1} \, \nabla_{\x_0} \mathcal{E}$ in \eqref{eqn:sensobs}. 
The inverse of the 4D-Var Hessian approximates the covariance matrix of the reanalysis error \cite{Gejadze_2008,Gejadze_2008b}. 
The solution $\mu_0$ will be referred to as ``supersensitivity'', and is a crucial ingredient for the computation of forecast sensitivities to all data assimilation parameters . The present work focuses on efficiently solving the linear system \eqref{eqn:linsys}, as it presents the main computational burden of the entire methodology.

\subsection{Forecast sensitivity to the 4D-Var parameters}

From \eqref{eqn:sensobs} the forecast sensitivity to observations is obtained as follows:
\begin{eqnarray*}
\mu_k &=& \M_{0,k}\,\mu_0\,, \\  
\nabla_{\y_k} \mathcal{E}(\xa_0) &=& \R_k^{-1} \, \HH_k \, \mu_k\,.
\end{eqnarray*}
The index $k$ selects the observation time $t_k$. The supersensitivity $\mu_0$ at $t_0$ is propagated forward to time $t_k$ using the tangent linear model, to obtain the vector $\mu_k$. This solution is applied the linearized observation operator $\HH_k$, and is scaled by $\R_k^{-1}$, the inverse covariance matrix of the observational errors. The sensitivity equations for other parameters can be found in \cite{Daescu_2008}. For example, the forecast sensitivity to the background estimate is
\[
\nabla_{\xb_0} \mathcal{E}(\xa_0) = \B_0^{-1} \, \mu_0\,. 
\]
This provides insight about the meaning of supersensitivity: it represents a time-dependent field that quantifies the sensitivity of the forecast score to the information assimilated at a certain time. At $t_0$ this information  is the background, and at other times is the observations.

\section{Numerical Tests with the Shallow Water Equations}\label{sec:swe}

\subsection{Numerical model}

The first model used to study the performance of the computational methodology is based on the shallow-water equations ({\sc swe}). The two-dimensional PDE system (\ref{swe}) approximates a thin layer of fluid inside a shallow basin:
\begin{eqnarray}
 \frac{\partial}{\partial t} h + \frac{\partial}{\partial x} (uh) + \frac{\partial}{\partial y} (vh) &=& 0 \nonumber \\
 \frac{\partial}{\partial t} (uh) + \frac{\partial}{\partial x} \left(u^2 h + \frac{1}{2} g h^2\right) + \frac{\partial}{\partial y} (u v h) &=& 0  \label{swe} \\
 \frac{\partial}{\partial t} (vh) + \frac{\partial}{\partial x} (u v h) + \frac{\partial}{\partial y} \left(v^2 h + \frac{1}{2} g h^2\right) &=& 0 \;.
\nonumber
\end{eqnarray}
Here $h(t,x,y)$ is the fluid layer thickness, and $u(t,x,y)$ and $v(t,x,y)$ are the components of the velocity field.  The gravitational acceleration is denoted by $g$. The spatial domain is $\Omega = [-3,\,3]^2$ (spatial units), and the integration window is $t_0 = 0 \le t \le t_\textrm{f} = 0.1$ (time units).

The numerical model uses a finite volume-type scheme for space discretization and a fourth-order Runge-Kutta scheme for time discretization \cite{Wendroff_1998}.  A square $q \times q$ discretization grid is used, and the numerical model has $n = 3 q^2$ variables 
\[ 
\x = \begin{bmatrix} \hat{h} \\ \hat{uh} \\ \hat{vh} \end{bmatrix} \in \mathbb{R}^{n} \;.
\]
We call the discretized system of equations {\em the forward model} ({\sc fwd}), used to simulate the evolution of the nonlinear system \eqref{swe} forward in time.  We are interested in computing the derivatives of a cost function $\Jfunc(\x_0)$ with respect to model parameters, like the initial condition. These derivatives can be computed efficiently using adjoint modeling. The theory and applications of adjoint models to data assimilation can be found in \cite{WangNavon, SanduZhang_2008}. The distinction is made between continuous adjoints, obtained by linearizing the differential equations, and discrete adjoints, obtained by linearizing the numerical method. Construction of adjoint models is a work intensive and error prone process. An attractive approach is automatic differentiation (AD) \cite{griewank1989automatic}. This procedure parses the source code of the {\sc fwd} model and generates the code for the discrete adjoint model using line by line differentiation. 

We build the adjoint {\sc swe} model through automatic differentiation using the TAMC tool \cite{giering1997tangent, TAMC_1998}. The tangent-linear model ({\sc tlm}) propagates perturbations forward in time. The first-order adjoint model ({\sc foa}) propagates perturbations backwards in time, and efficiently computes the gradient of a scalar cost function of interest ($\nabla_{\x_0} \Jfunc$). The second-order adjoint model ({\sc soa}) computes the product between the Hessian of the cost function and a user-defined vector ($\nabla^2_{\x_0,\x_0} \Jfunc \cdot u$) \cite{Cioaca_2011}. Second-order adjoint models are considered to be the best approach to compute Hessian-vector products, but have yet to become popular in practice because of their computational demands. When one does not have access to the second-order adjoint, Hessian-vector products can be computed through various approximations, such as finite differences of first order adjoints. 

The overhead of running adjoint models has to be taken into account for the design of the computational strategy. 
Table \ref{Table:CPUTimes_exp} presents the CPU times of {\sc tlm}, {\sc foa} and {\sc soa} shallow models, normalized with respect to the CPU time of a single {\sc fwd} model run. One {\sc soa} integration is about $3.5$ times more expensive than a single first-order adjoint run, while the {\sc foa} takes $3.7$ times longer than the forward run. The adjoint model runs take a significant computational time. This effort depends on the numerical methods used in the {\sc fwd} model, and on the automatic differentiation tool employed. For certain numerical methods  it is possible to develop efficient 
strategies based on reusing computations, which lead to adjoint times smaller than forward model times. An example can be found in \cite{Cioaca_2011} where the adjoint {\sc swe} equations  are derived by hand and then solved numerically.

\begin{table}
{
\centering
\begin{tabular}{|c|c||c|c|}
  \hline
  {\sc fwd} & $1$ & & \\
 \hline
  {\sc tlm} & $2.5$ & {\sc fwd} + {\sc tlm} & $3.5$ \\
 \hline
  {\sc foa} & $3.7$ & {\sc fwd} + {\sc foa} & $4.7$\\
 \hline
  {\sc soa} & $12.8$ & {\sc fwd} + {\sc tlm} + {\sc foa} + {\sc soa} & $20$ \\
 \hline
\end{tabular}
\caption{Normalized CPU times of different sensitivity models. The forward model takes one time unit to run.}
\label{Table:CPUTimes_exp}
}
\end{table}

\subsection{Data Assimilation Scenario}\label{sec:das}

The 4D-Var data assimilation system used in the numerical experiments is set up as follows:
\begin{itemize}
\item The computational grid uses $q=40$ grid points in each directions, for a total of $4800$ model variables.
 The timestep is $0.001$ (time units).
\item The reference solution is obtained as follows. The initial $h$ field is a Gaussian bell 
centered on the grid. The initial $u$ and $v$ are constant fields. We run the forecast model from the initial solution for $100$ time steps. The solutionprovides the reference trajectory for the experimental setup. 
\item The background solution $\x^b$ is generated by adding a correlated perturbation to the reference solution $\x=[h,u,v]$.
The background error covariance $\B_0$ corresponds to a standard deviation of $5\%$ of the reference field values.
The spatial error correlation uses a Gaussian decay model, with a correlation distance of $5$ grid points. This dictates how the 4D-Var method spreads the information from one grid point to its neighbors.
\item Synthetic observations are generated from the reference model results. The observation frequency is set to once every $20$ time steps. We add normal random perturbations to simulate observation errors.
The observation error covariance matrix $\R$ is diagonal (i.e., the observation errors are uncorrelated). The standard deviation is $1\%$ of the largest absolute value of the observations for each variable.
\item The observation operator $\Hobs$ is linear and selects observed variables at specified grid points. 
\end{itemize}
We use the L-BFGS-B solver \cite{zhu1997algorithm} to minimize the 4D-Var cost function. We allow the solver to run for $400$ iterations (which reduces the norm of gradient of the 4D-Var cost function from a magnitude of $1e+7$ to $1e-4$). Note that one cannot afford to obtain such a high quality optimal solution with a large-scale model. The {\sc swe} test case allows to compute the sensitivity to observations in a setting where numerical optimization errors are negligible.  

\subsection{Particularities of the linear system}\label{sec:ls}

The solution of the linear system \eqref{eqn:linsys} is the central step of the entire computational process. As mentioned in Section \ref{sub:rhs}, the right hand side is the gradient of the forecast aspect with respect to initial conditions, and is obtained at the cost of one {\sc foa} run. The adjoint model propagates backward in time the mismatch between the forecast and the verification.

The system matrix in \eqref{eqn:linsys} is the Hessian of the 4D-Var cost function, evaluated at the reanalysis. For large-scale models like the atmosphere, the Hessian cannot be computed and manipulated in an explicit form due to its dimension. In practice, one evaluates directly the Hessian-vector product by running the second-order adjoint model. When {\sc soa} is not available, one can approximate Hessian-vector products through finite differences of {\sc foa} gradients. 
\begin{equation}
 \nabla_{\x_0,\x_0}^2 \mathcal \Jfunc(\xa_0) \cdot \u \approx \frac{\nabla_{\x_0} \mathcal \Jfunc(\xa_0 + \epsilon \cdot \u)^T - \nabla_{\x_0} \mathcal \Jfunc(\xa_0)^T}{\epsilon}\,.
\end{equation}
A third method to compute Hessian-vector products is the Gauss-Newton approximation of the Hessian, also known in literature as the ``Hessian of the auxiliary cost function'':
\begin{equation}
 \nabla_{\x_0,\x_0}^2 \mathcal \Jfunc(\xa_0) \cdot \u \approx \B_0^{-1}  \cdot \u + \sum_{k=1}^{N}  \M_{0,k}^T \HH_k^T \R_k^{-1}  \HH_k\, \M_{0,k}\cdot u\,.
\end{equation}

The formulation above is obtained in a similar fashion to the formulation of incremental 4D-Var \cite{Courtier_1994}, by differentiating the 4D-Var cost function and ignoring higher-order terms. These higher-order terms are negligible when the solution is close to the optimum. Computationally, the Gauss-Newton Hessian-vector product is obtained by running the {\sc tlm} model forward in time starting from the seed vector, and then using its output to initialize a {\sc foa} model run backward in time. 

For our {\sc swe} model, both finite difference and Gauss-Newton approximations provide Hessian-vector products that verify within machine precision with the Hessian-vector products obtained from second-order adjoint models. However, finite difference is less stable than Gauss-Newton since it relies on perturbing the system. 

Yet another strategy is to build limited-memory approximations of the Hessian from information collected during the data assimilation process. In \cite{Tremolet_2007} the authors use the Lanczos pairs generated by the iterative solver employed to minimize the 4D-Var cost function. This type of approximation is usually helpful for building preconditioners, but is not accurate enough to be used as the system matrix in \eqref{eqn:linsys}.

Corresponding to the spatial discretization chosen for our experiment, the size of the model solution is $4800$ variables. Accordingly, the size of the 4D-Var Hessian matrix is $4800 \times 4800$. The explicit form of this matrix can be obtained through matrix-vector products with the $e_i$ unity vectors ({\sc soa} model). This is not feasible in practice, but our SWE model is small enough to allow us to build the full Hessian and analyze its properties. Thus, we find out the Hessian is symmetric to machine precision, which confirms the superior quality of second-order information obtained with the {\sc soa} model. Also, because the 4D-Var optimization problem in Section \ref{sec:das} is solved accurately, the reanalysis is close to the optimum and the 4D-Var Hessian evaluated at this point is positive definite. Our tests show that when evaluated far from the optimum, the 4D-Var Hessian is indefinite. This has consequences for real-time operations where only a limited number of iterations are allowed. 


The structure of the Hessian matrix exhibits some regularities, characteristic to information matrices and their covariance counterparts. In literature, this structure is known as ``near block-Toeplitz'' \cite{Toeplitz}. The first $1600$ rows correspond to the model variables of $h$, the next $1600$ rows to $u$ and the last $1600$ to $v$. The matrix elements scale differently in each one of these three blocks. Some obvious features occur on the diagonals, rows and columns, spaced every 40 or 1600 rows and columns. This hints at the fact that the 4D-Var Hessian approximates the inverse of the covariance matrix of the reanalysis errors \cite{Gejadze_2008,Gejadze_2008b}. We interpret these patterns as arising from due to the discretization scheme stencil (each point of the grid is correlated to its East, West, North, and South neighbors). In addition, each variable is weakly connected to the other two variables, corresponding to a distance of 1600 rows/columns. This structure can be predicted without building 
the explicit form of the Hessian, from prior information such as the background error covariance matrix $\B_0$.

\begin{figure}[ht]
\centering
\includegraphics[width=8cm]{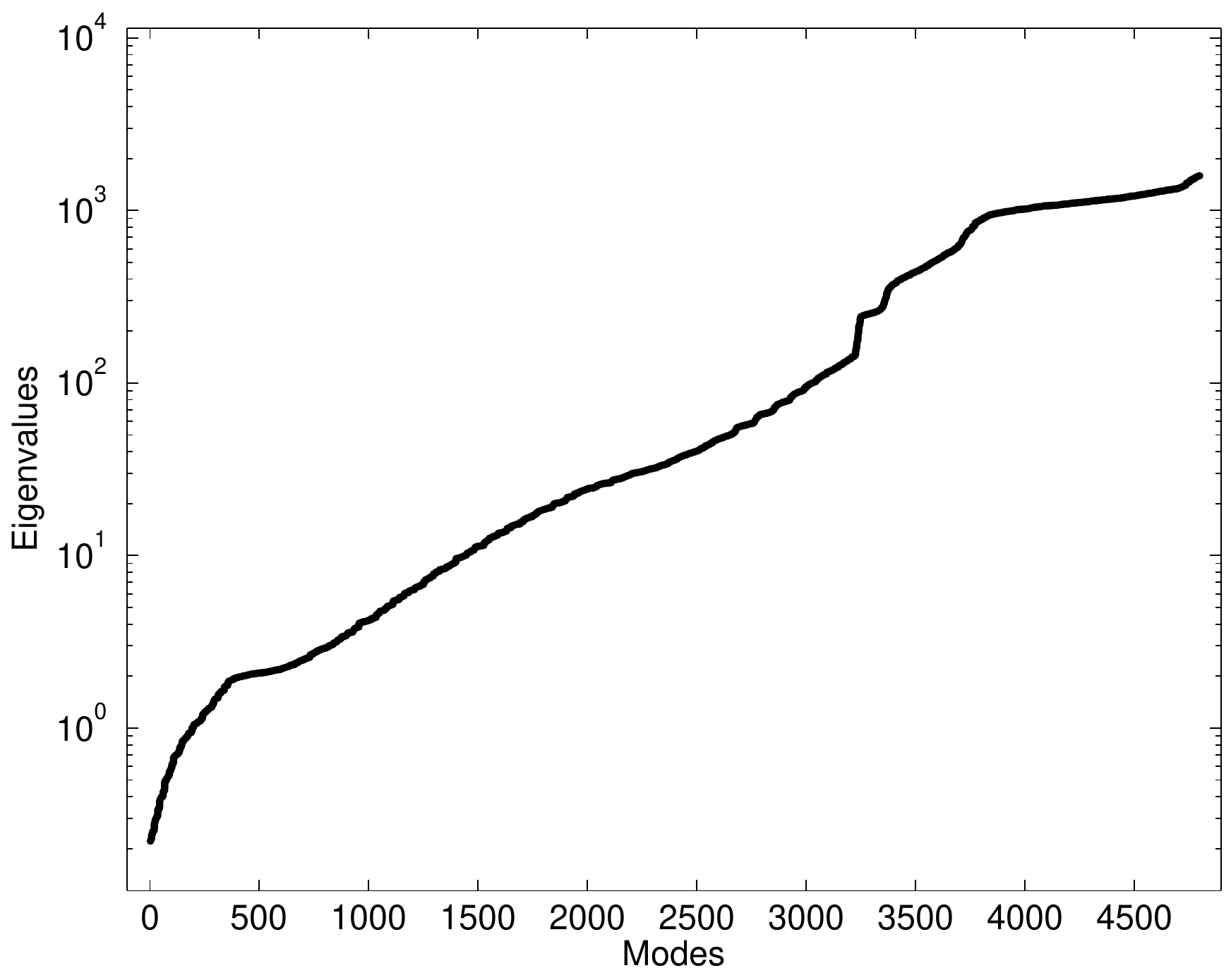}
\caption{ Eigenvalues of the {\sc swe} 4D-Var Hessian at the reanalysis (optimal solution), sorted in ascending order. }
\label{fig:eigvals}
\end{figure}

The spectrum of the matrix is of great interest for our analysis, since it will influence the convergence of the iterative solvers. The eigenvalues of the {\sc swe} Hessian are displayed in Figure \ref{fig:eigvals}, sorted in ascending order. The condition number of the Hessian (ratio between largest and smallest eigenvalues) is $\sim 10^4$, which makes the matrix moderately well-conditioned. However, since the eigenvalues are not clustered together, we expect slow convergence.

\subsection{Matrix-free linear solvers}

The choice of solvers for the linear system  \eqref{eqn:linsys} is limited to  ``matrix-free'' algorithms. Direct solvers and basic iterative methods are ruled out since they require the full system matrix, which is not available. Krylov-based iterative solvers require only matrix-vector products and exhibit superior performance over basic iterative methods. However, their convergence depends on the eigenvalues of the system matrix. As seen in Figure \eqref{fig:eigvals}, the Hessian is positive definite, but its spectrum is scattered. Preconditioning can considerably improve the convergence of iterative solvers.

Additional challenges arise in large-scale 4D-Var data assimilation. The reanalysis can be far from the minimizer, when the minimizing algorithm is stopped before reaching the minimum; in this case, the resulting Hessian matrix can be indefinite. Although by definition a Hessian matrix is symmetric, the symmetry can be lost when approximations such as finite differences are employed. In an operational setting where the sensitivities are used to target adaptive observations, results have to be delivered in real time; the key is to provide the best possible solution in a given time.


The matrix-free iterative solvers used to solve the {\sc swe} supersensitivity system \eqref{eqn:linsys} are listed in Table \ref{tab:kry}. The list includes the most popular algorithms currently used for large linear systems. Detailed information about each solver can be found in the scientific literature \cite{Vorst,Saad}.
\begin{table}[ht]
\centering
\begin{tabular}{|c|c|c|}
\hline
Generalized Minimum Residual & GMRES    & nonsymmetric  \\
\hline
Minimal Residual             & MINRES   & symmetric \\
\hline
Conjugate Gradients          & CG       & symmetric positive-def. \\
\hline
Quasi-Minimum Residual       & QMR      & nonsymmetric \\
\hline
Biconjugate Gradients Stabilized & BiCGSTAB & nonsymmetric \\
\hline
Conjugate Gradients Squared  & CGS      & nonsymmetric \\
\hline
Least Squares                & LSQR     & nonsymmetric \\
\hline
\end{tabular}
\caption{ List of iterative methods used to solve the ({\sc swe}) system \eqref{eqn:linsys}. }
\label{tab:kry}
\end{table}

We used the iterative solvers implemented in the PETSc \cite{PETSc} software package. PETSc supports matrix and vector operations and contains an extensive set of solvers and preconditioners. We interfaced PETSc with our shallow water model and solved the linear system with each of the methods above. Also, we double-checked the results with our own Fortran and MATLAB implementation of the algorithms. The initial guess was set to a vector of zeroes and no preconditioner was used for the results presented in this section. We compare the convergence of the linear solvers by monitoring the decrease in the residual norm and the error norm at each iteration. The error norm was computed as a root mean square error with respect to a reference solution $\mu_0^{REF}$ obtained by solving the system directly using the full Hessian, and this error metric has the following expression:

\begin{equation}
  RMSE = \frac{\Vert\mu_0 - \mu_0^{REF}\Vert}{\sqrt{n}}.  
\end{equation}

\begin{figure}[ht]
\centering
\subfigure[Error norm]{
\includegraphics[height=0.5\textwidth,width=0.7\textwidth]{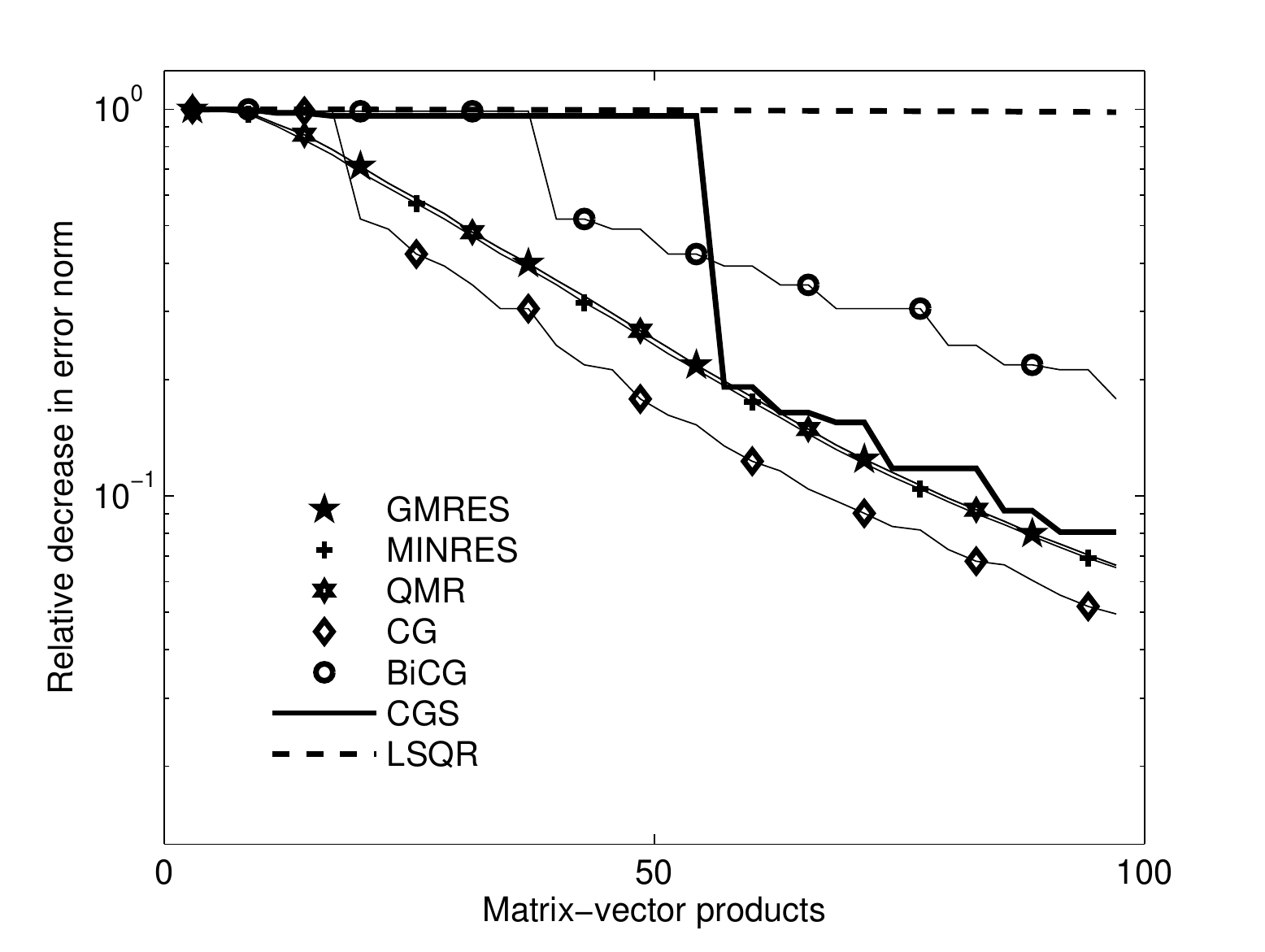}
\label{fig:relerr}
}
\subfigure[Residual norm]{
\includegraphics[height=0.5\textwidth,width=0.7\textwidth]{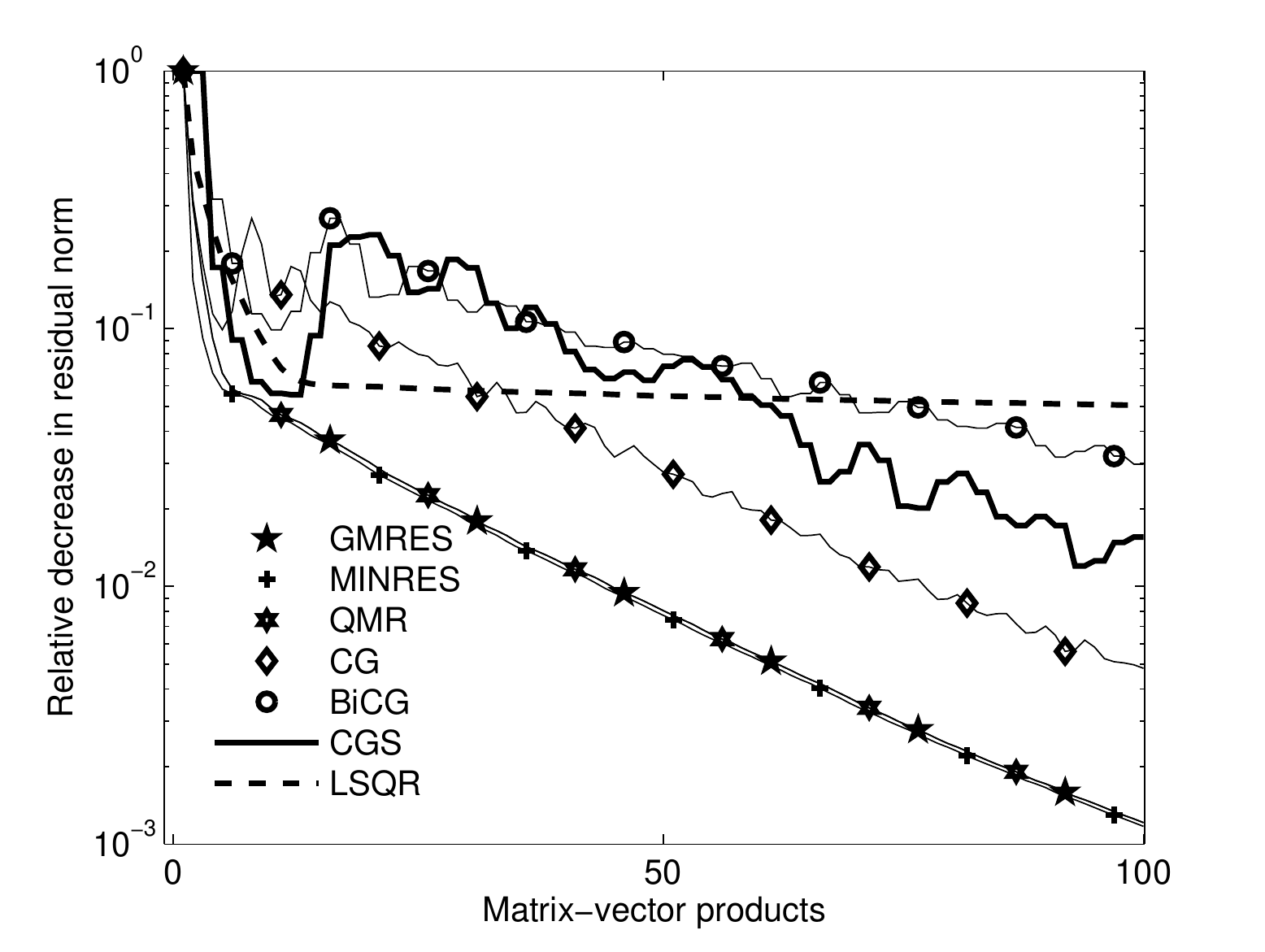}
}
\caption{Convergence of non-preconditioned iterative solvers for the ({\sc swe}) supersensitivity system  \eqref{eqn:linsys}. }
\label{fig:swe-convergence}
\end{figure}

We allocate a budget of $100$ matrix-vector products as {\sc SOA} runs. BiCGSTAB, CGS use $2$ matrix-vector products per iteration, which means $50$ iterations. The other solvers use just $1$, so they will run for $100$ iterations within our budget. Figure \ref{fig:swe-convergence}(a) plots the relative decrease in the norm of the error and Figure \ref{fig:swe-convergence}(b) the relative decrease in the norm of the residual. Table \ref{table:linsys} presents the solution error and residual norm decrease after 100 matrix-vector products of each solver.

\begin{table}
{
\centering
\begin{tabular}{|c|c|c|}
\hline
Solver   & Relative decrease & Relative decrease \\
         &  in residual norm &  in error norm    \\
\hline \hline
GMRES    & $2.219$e-1          & $6.62$e-2          \\
\hline
MINRES   & $2.164$e-1          & $6.53$e-2          \\
\hline
PCG      & $9.461$e-1          & $4.95$e-2          \\
\hline
QMR      & $2.219$e-1          & $6.62$e-2          \\
\hline
BiCGSTAB & $9.461$e-1          & $5.54$e-2          \\
\hline
CGS      & $1.124$e-1          & $1.48$e-2          \\
\hline
LSQR     & $9.792$e0           & $9.83$e-1          \\
 \hline
\end{tabular}
\caption{ Solution error and residual norms after 100 matrix-vector products of each solver for the ({\sc swe}) supersensitivity system \eqref{eqn:linsys}.
The scaling is done with respect to the initial guess error and residual norms, respectively.  }
\label{table:linsys}
}
\end{table}

The decrease in the solution error and residual norms are as expected from the theory of Krylov solvers. CG provides the best error reduction. GMRES, MINRES and QMR show the best performance for reducing the residual. CG is known for its superior performance over other solvers when dealing with symmetric and positive definite matrices. It acts on reducing the A-norm of the error, as opposed to GMRES, MINRES and QMR, which act upon the residual. For symmetric positive definite matrices, the latter three are equivalent, which explains their similar behavior. CGS and BiCGSTAB exhibit a slow initial convergence, but CGS eventually catches up with GMRES. LSQR has the worst performance, confirming that a least-squares approach is not suitable for solving this problem.
In consequence, CG would be the ideal solver to use when we can guarantee the system matrix is symmetric and positive-definite. Otherwise, one should use GMRES (or MINRES), with the amendment that the numerical workload per iteration is slightly larger than for CG. 


\subsection{Preconditioned Krylov solvers}

We next explore preconditioning strategies to improve the convergence of the iterative methods. The Krylov solvers perform better when the matrix eigenvalues are clustered. As seen in Figure \ref{fig:eigvals}, the eigenvalues of the {\sc swe} Hessian matrix are scattered across various orders of magnitude. 
This explains why no method converged to the actual solution. 


Building effective preconditioners for the supersensitivity linear system \eqref{eqn:linsys} is challenging. Preconditioners require a good understanding of the underlying problem and the structure of the matrix; this is difficult without having access to the full system matrix. The matrix-free constraint excludes certain preconditioning techniques such as incomplete factorizations, wavelet-based, or variations of the Schur complement. Moreover, basic preconditioners such as diagonal cannot be constructed solely from matrix-vector products, without a significant computational effort. We consider here preconditioning strategies that rely on curvature information collected during the numerical minimization process. Predicting the structure of the Hessian matrix can also help with the solution of the problem. 
We next describe the proposed preconditioners.

\subsubsection{Diagonal of Hessian}

The diagonal of the matrix is one of the most popular practical preconditioners, and was proved to be the optimal diagonal preconditioner in \cite{Strauss_1955}. When we only have access to the matrix under the form of an operator, its diagonal is not readily available. Therefore, we use the diagonal preconditioner in this test only as a reference for the performance of the other preconditioners. In a real setting, one has access to neither the actual diagonal, nor banded or arrow preconditioners.

\subsubsection{Diagonal of the background covariance matrix}

Preconditioners that do not require any supplementary computations can be obtained from $\B_0$, the covariance matrix of the background errors in 4D-Var. In practice, this matrix cannot be manipulated with ease due to its size. However, its diagonal is accessible, and we use it as a preconditioner in the following tests. This choice has been reported to provide better convergence in incremental 4D-Var under certain conditions \cite{Tremolet_2007}. 

\subsubsection{Row sum}

The system matrix \eqref{eqn:linsys} approximates the inverse of a covariance matrix. Covariance matrices have their larger elements on the diagonal, and under some conditions they have a diagonally dominant structure. Consequently we use the sum of row elements to build an approximation of the diagonal. This can be computed with just one second-order adjoint run, where the Hessian is multiplied by a vector of ones. The diagonal preconditioner used in our tests is built from the output of the second-order adjoint and taking the absolute value.

\subsubsection{Probing and extrapolating}



This approach takes advantage of the results in \cite{Zupanski_1993,Navon_1996} where the possibility of block diagonal approximations of the 4D-Var Hessian is explored. The values for a certain variable and for a certain vertical level (not applicable here since we have a 2D model) are assigned a constant value. We approximate these values by using Hessian-vector products to ``probe'' the matrix. For our three-variable model we run three Hessian-vector products with unity vectors to extract one column (row) of the Hessian at one time. The value of the corresponding diagonal element is used as an approximation for all diagonal elements in that block. 

To be specific, we consider three unity vectors for our $4800\times4800$ Hessian that have the value $1$ at positions $1$, $1601$ and $3201$ respectively, and zeros everywhere else. The corresponding Hessian-vector products will extract the columns $1$, $1601$ and $3201$, which correspond to the three different variables in our Hessian. The approximation uses the value found at coordinates $(1,1)$ for the entire first diagonal block (up to coordinates $1600$, $1600$), the value found at coordinates $(1601,1601)$ for the entire second block, and so forth. This approximation can be refined by probing for more elements from the same block. If there are many blocks that have to be probed and the computational burden increases significantly, one can employ coloring techniques to probe for more than one element with the same matrix-vector product. 

\subsubsection{Quasi-Newton approximation}

The Hessian matrix can also be approximated from data collected throughout the minimization process. Quasi-Newton solvers such as L-BFGS build Hessian approximations, and refine them with information generated at each iteration. These approximations are sufficiently accurate along the descent directions to improve the convergence of the minimization iterations. The approximations preserve matrix properties such as symmetry and positive definiteness, and allow limited memory implementations appropriate for large-scale models. We store the approximation of the Hessian as generated over the last 10 iterations of minimizing the 4D-Var cost function with L-BFGS. This will be used as a preconditioner for the linear system and does not require any supplementary model runs. Our tests showed that using more than 10 vector pairs does not improve further the quality of the resulting preconditioner. 

\subsubsection{Eigenpairs}

This preconditioning method is borrowed from 4D-Var data assimilation literature \cite{Tremolet_2007}. During the minimization of the 4D-Var cost function the leading eigenvalues and eigenvectors are calculated via a Lanczos process. An approximation of the Hessian (evaluated at the current reanalysis) can be generated from the leading eigenvalues or eigenvectors, and used as a preconditioner for the supersensitivity system \eqref{eqn:linsys}. In our tests we use the leading 50 eigenpairs to approximate the Hessian.

\subsubsection{Randomized SVD}

Randomized SVD \cite{RandSVD_2007} computes an approximate singular value decomposition of a matrix only available as an operator. The algorithm requires two ensembles of matrix-vector products, and one singular value decomposition and one QR decomposition with smaller matrices. All matrix-vector products can be executed in parallel as they are independent of each other. The number of input vectors used can vary and the accuracy of the approximation is proportional to the size of the ensemble. For our tests we used 50 different input vectors.

\begin{table}[ht]
\centering
\begin{tabular}{|c||c|c|c|c|c|c|c|}
\hline
Preconditioner & Relative decrease & Relative decrease \\
               & in residual norm  & in error norm \\
\hline\hline
None           & $1.3$e-3         & $7.2$e-3 \\         
\hline
Diagonal       & $8.0$e-5         & $1.2$e-3 \\
\hline
Coloring       & $8.0$e-5         & $1.2$e-3 \\
\hline
Row sum        & $1.2$e-4         & $1.9$e-3 \\
\hline
L-BFGS         & $3.8$e-4         & $1.6$e-2 \\
\hline
Eigenpairs     & $8.0$e-5         & $1.7$e-3 \\
\hline
RandSVD        & $8.0$e-5         & $1.2$e-3 \\
\hline
\end{tabular}
\caption{ Solution error and residual norms after 100 non-preconditioned iterations of GMRES for the ({\sc swe}) supersensitivity system \eqref{eqn:linsys}.
The scaling is done with respect to the initial guess error and residual norms, respectively. }
\label{Table:precond}
\end{table}

\subsubsection{Performance of preconditioned algorithms}

\begin{figure}[ht]
\centering
\subfigure[Error norm]{
\includegraphics[height=0.5\textwidth,width=0.7\textwidth]{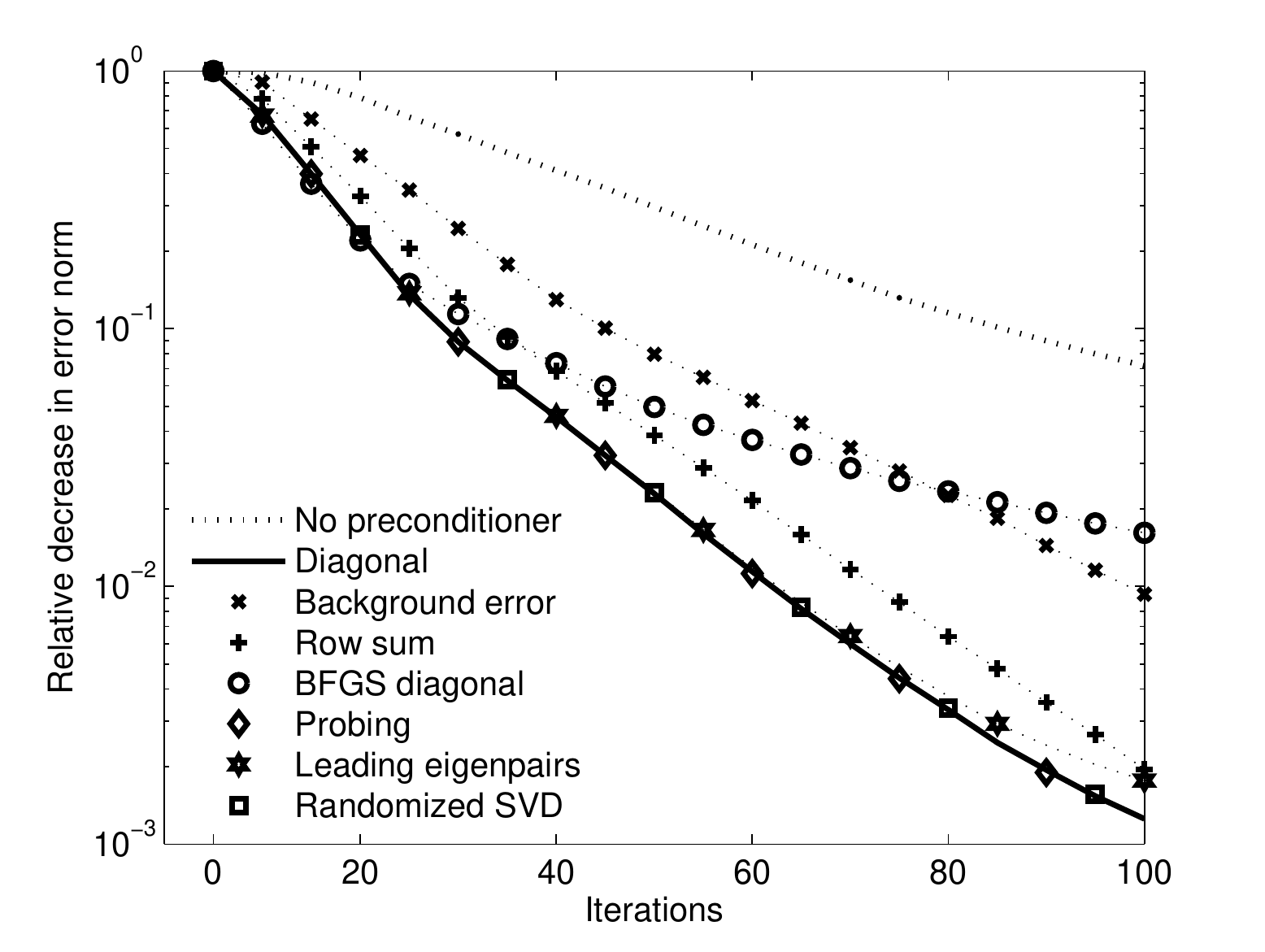}
\label{fig:sweprecerr}
}
\subfigure[Residual norm]{
\includegraphics[height=0.5\textwidth,width=0.7\textwidth]{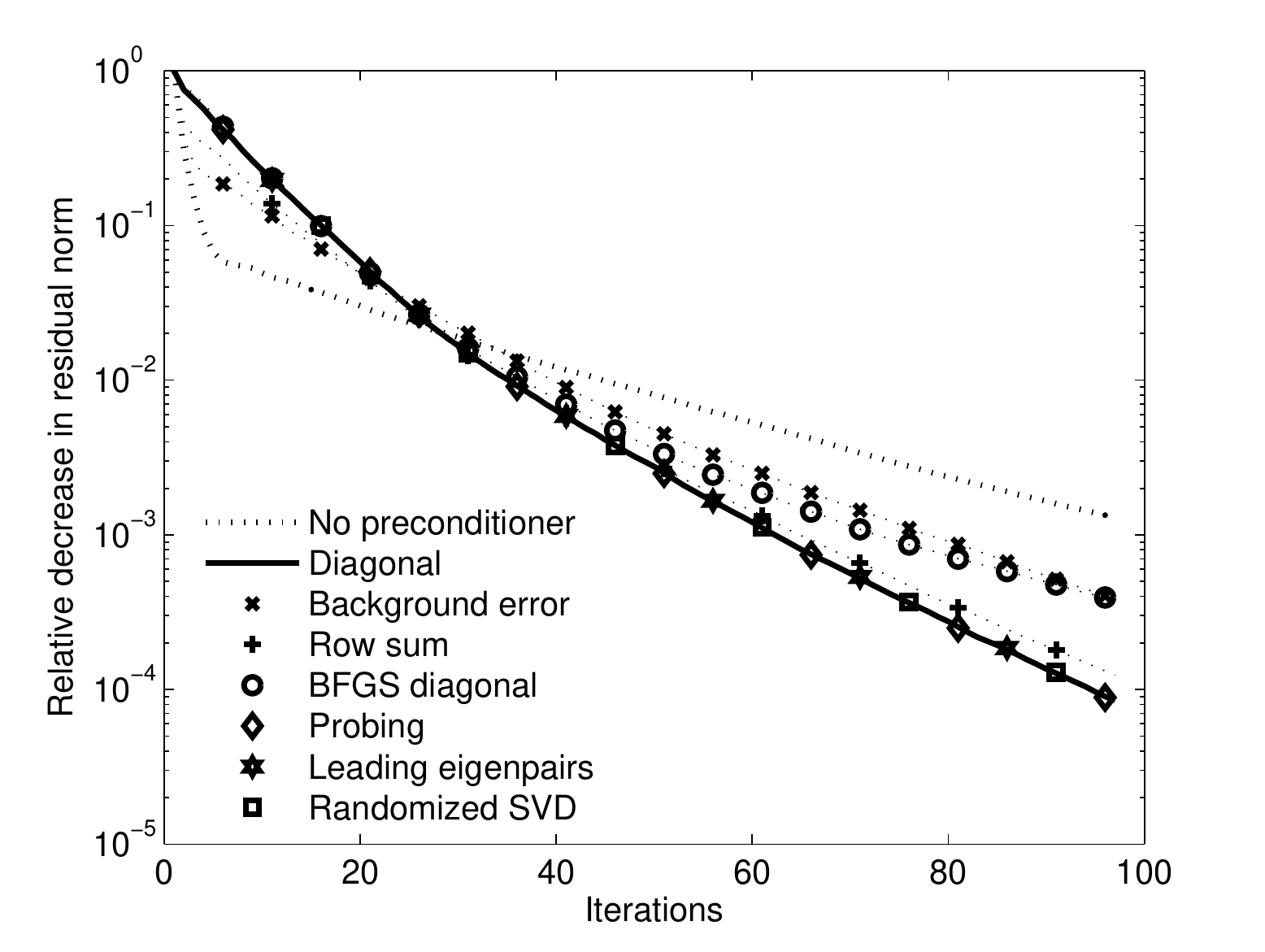}
}
\caption{ Convergence of non-preconditioned iterative solvers for the ({\sc swe}) supersensitivity system \eqref{eqn:linsys}. }
\label{fig:sweprecres}
\end{figure}

%

The experiments to compare the performance of the preconditioners were conducted with GMRES as the linear solver, because of its generality. The norm of the error against the reference solution and that of the residual are shown in Table \ref{Table:precond} and Figures \ref{fig:sweprecerr}, \ref{fig:sweprecres}. A comparison with the results in Table \ref{tab:kry} and Figures \ref{fig:relerr}, \ref{fig:swe-convergence} reveals that all preconditioners improve convergence. L-BFGS LMP starts off with the best decrease, but then it stops accelerating, and after 100 iterations has the worst performance among all preconditioners. The preconditioners formed from probing, leading eigenpairs, and randomized SVD, perform almost as well as the exact diagonal. Finally, the row sum preconditioner also shows good results, comparable to the latter preconditioners.

The conclusion is that some preconditioners can decrease the error after $100$ iterations by a factor of up to $100$. 
After $25$ iterations the preconditioned algorithm reaches the same accuracy that the unpreconditioned algorithm achieves after $100$ iterations.
This improvement of $75$\% in the computation time is very significant for large-scale models.

\subsection{Multigrid solver}

Multigrid (MG) describes a class of numerical methods that speed up numerical solutions by alternating computations on coarser or finer levels \cite{Fedorenko_1961, Hackbusch_2003}. These methods can be defined geometrically (using a grid) or purely algebraically. We refer to each fine-grid-to-coarse-grid sweep as a ``multigrid cycle'', ``V-cycle'' or ``cycle'' for short. 


Our linear system \eqref{eqn:linsys} is appropriate for the multigrid approach because one can run the {\sc swe} model on different spatial discretizations. Consider the $40 \times 40$ grid used in the previous tests as the fine-level grid ($4800$ variables). We can simulate the same scenario coarser grid, for example $20 \times 20$ ($1200$ variables) and $10 \times 10$ (300 variables). For simplicity and clarity, we  use only the first two levels in our test. 

Traditional MG uses smoothers that require the full matrix, and one challenge is to build a matrix-free approach.  Here we use GMRES as smoother. The MG theory does not guarantee convergence for Krylov-based methods, but there are reports of them being used successfully. 
A second challenge consists in designing the operators that transfer the problem between grids. One needs to restrict the residual of the linear system from the fine grid to the coarse grid and to prolongate the correction from the coarse grid back to the fine grid. 
We use a projection operator that computes the mean value of a square of size $2 \times 2$ to reduce our field by a factor of four;
the interpolation operator is the transpose of the projection operator.


To assess the performance of the two-level multigrid method we limit the number of model runs to 100. We run multigrid GMRES with one, two and three cycles, and allocate the $100$ model runs uniformly across cycles and levels. For MG with one cycle we allocate the model runs as 
33 model runs to the initial fine grid smoother ($F$), then 33 model runs to the coarse grid solver ($C$), and 34 model runs on the final fine grid smoothing. For two cycle we distribute these 100 model runs as $20F + 20C + 20F + 20C + 20F$. The same applies for three cycles, where we have $~14$ model runs on each grid. We are interested in a conclusive reduction in the residual (or error), especially after projecting the correction from the coarse grid to the fine grid. 

Table \ref{tab:multigrid} shows theMG solver results. The rows represent the different MG scenarios described above, plus a standard approach without MG, on the first line. The columns represent MG cycles. Each cycle is composed of two levels: fine and coarse. The MG algorithm starts on the fine grid by smoothing out the errors, then projects the residual of the intermediate solution on the coarse grid, and performs another smoothing of the errors. The result is projected back to the fine grid and used to correct the solution. This is called ``Correction Scheme'' as opposite to ``Full Approximation Scheme'' and is repeated for as many cycles as necessary. In each table entry we display the residual and error norms. For fine grid columns, the norms are computed on the fine grid, and correspond to the solution obtained after smoothing. For coarse grid columns, the displayed norms were still computed on the fine grid, after prolongating the correction from the coarse grid to the fine grid, and applying it to 
the solution.  We show all the intermediate solutions in order to analyze the MG 
behavior for each cycle. The solution error norm decreases after projecting and applying the correction from the coarse grid to the fine grid after each stage. This was not trivial to accomplish, as it required crafting the prolongation operator as described above. The improvement is not reflected by the solution residual norm which sometimes shows an increase after prolongation, for example when using MG with one cycle. By comparing the final solution error norm obtained for different MG scenarios, it is inferred that better results are obtained with using fewer cycles, and more smoother iterations per cycle. This can be explained in terms of the Krylov solvers having more iterations available to build the Krylov space; the Krylov space information is lost when switching from one grid to another.

\begin{table}[ht]
\centering
\begin{tabular}{|c||c|c|c|c|c|c|c|}
  \hline
  Cycle    &   1     &    1    &   2     &    2    &  3     &    3     & Final   \\
  \hline
  Level    & Fine    & Coarse  & Fine    & Coarse  & Fine    & Coarse  & Fine    \\
 \hline
 \hline
  Residual & & & & & &                                                 & $4.0$.e-4 \\
  Error    & & & & & &                                                 & $1.9$e-2 \\
 \hline
  Residual & $1.1$e-2 & $3.1$e-2 & & & &                               & $7.0$.e-4 \\
  Error    & $7.7$e-2 & $4.2$e-2 & & & &                               & $2.6$e-2 \\
 \hline
 \hline
  Residual & $1.1$e-2 & $1.4$e-1 & $3.0$e-3 & $8.1$e-2 & &             & $1.0$e-3 \\
  Error    & $9.1$e-2 & $6.4$e-2 & $5.5$e-2 & $4.4$e-2 & &             & $4.0$e-2 \\
 \hline
 \hline
  Residual & $2.5$e-2 & $3.9$e-1 & $1.1$e-2 & $2.7$e-1 & $8.0$e-3 & $1.8$e-2 & $6.0$e-3   \\
  Error    & $1.1$e-2 & $8.3$e-2 & $6.7$e-2 & $6.5$e-2 & $5.3$e-2 & $5.2$e-2 & $4.4$e-2 \\
 \hline

\end{tabular}
\caption{ Residual and error norms of solutions obtained at each multigrid stage ({\sc swe}). }
\label{tab:multigrid}
\end{table}

MG provides the ability to run the model at a coarser resolution which in turn reduces computing time. This is very useful when dealing with large-scale models and their adjoints. The results reported in Table \ref{tab:multigrid} are very good, even if they were produced using a basic MG algorithm. The performance of MG could be improved considerably by tuning the selection of coarse grids, building more accurate transfer operators, and testing additional matrix-free smoothers. 

\section{Numerical Tests with the Weather Research and Forecast Model}\label{sec:wrf}


In this section we consider a realistic test case based on the Weather Research and Forecasting (WRF) model. 

\subsection{Numerical model}

The WRF model \cite{WRF_2008} is a state-of-the-art numerical weather prediction system that can be used for both operational forecasting and atmospheric research. WRF is the result of a multiagency and university effort to build a highly parallelizable code that can run across scales ranging from large-eddy to global simulations. 
WRF accounts for multiple physical processes and includes cloud parameterization, landsurface models, atmosphere-ocean coupling, and broad radiation models. The terrain resolution can be as fine as 30 seconds of a degree. 

The auxiliary software package WRFPLUS \cite{xiao2005development} provides the corresponding tangent-linear and first-order adjoint models. WRFPLUS is becoming a standard tool for applications such as data assimilation \cite{schwartz2012impact} and sensitivity analysis \cite{CioacaSC_2011}. However, the adjoint model is work in progress and misses certain atmospheric processes. Because of this incompleteness, the computed sensitivities are only approximations of the full WRF gradients and Hessians. 
This will not affect the main conclusion of this study, namely that the proposed systematic approach to solving sensitivities to observations is feasible in the context of a real atmospheric model. Nevertheless, we expect that the sensitivity approximations have a negative impact on the convergence of the iterative solvers.

There is no second-order adjoint model developed for WRF to this point. This poses a challenge to our methodology, as it requires second-order derivatives. We consider several ways to approximate second-order information using the available tangent-linear or first-order adjoint models. First, we compute Hessian-vector products through finite differences of gradients obtained via first-order adjoint model. Unfortunately, our tests show that this approximation is marred by large errors and fails to produce useful results. Further investigation revealed that the adjoint model dampens the perturbations introduced in the system. The second approach is the Gauss-Newton approximation discussed in Section \ref{sec:ls}. The seed vector provides the initial condition to the tangent linear model, which propagates it to the final time. The result is mapped back to the initial time through the adjoint model. This is feasible for WRF since the required numerical tools are available. The Gauss-Newton approach introduces 
additional approximation errors in the second order sensitivity, beyond the incompleteness of the first order adjoint model.

WRF has the ability to perform forecasts on mesoscale domains defined and configured by the user. The simulation scenario selected  covers a region across the East Coast of North America, centered on Virginia, and takes place over a time period of $6$ hours starting on June 6th 2006 12:00 UTC. For simplicity, we assimilate only surface observations at the final time $t_0 + 6h$ obtained from NCEP. We start our simulations from reanalyzed fields, that is, simulated atmospheric states reconciled with observations (i.e., using data assimilation). In particular, we use the North American Regional Reanalysis (NARR) data set that covers the North American continent (160W-20W; 10N-80N) with a spatial resolution of 10 minutes of a degree, 29 pressure levels (1000-100 hPa, excluding the surface), a temporal resolution of three hours, and runs from 1979 until present. 

The spatial discretization is a regular grid with $30$ points on the East-West and North-South directions, and a horizontal resolution of 25 km. Since the atmosphere has different physical properties along with altitude, the vertical discretization involves 32 levels. A fixed time step of $30$ seconds is used. The wall clock time for one time step of the forward (WRF) model is $\sim 1.5$ seconds. The wall clock time for one time step of the adjoint (WRFPLUS) model is $\sim 4.5$ seconds, about three times larger. For finer grid resolutions or for nested grids the computational effort can increase significantly; one needs the power of parallel architectures for computing sensitivities in an operational setting.

The experiment starts with minimizing the 4D-Var cost function until the norm of the gradient is reduced from $\sim10^3$ to $\sim 10^{-3}$. The data assimilation procedure in WRFDA is an incremental approach revolving around the solution of a linear system as obtained with CG. The forecast error is obtained by comparing this reanalysis against a verification forecast represented by the corresponding NARR reanalysis. This forecast error was propagated backward in time through the adjoint model to obtain the right-hand side of the supersensitivity system \eqref{eqn:linsys}. All results below use Hessian-vector products computed using the Gauss-Newton approximation.

\subsection{Solution of the linear system}

\begin{figure}[ht]
\centering
\subfigure{
\includegraphics[height=0.5\textwidth,width=0.7\textwidth]{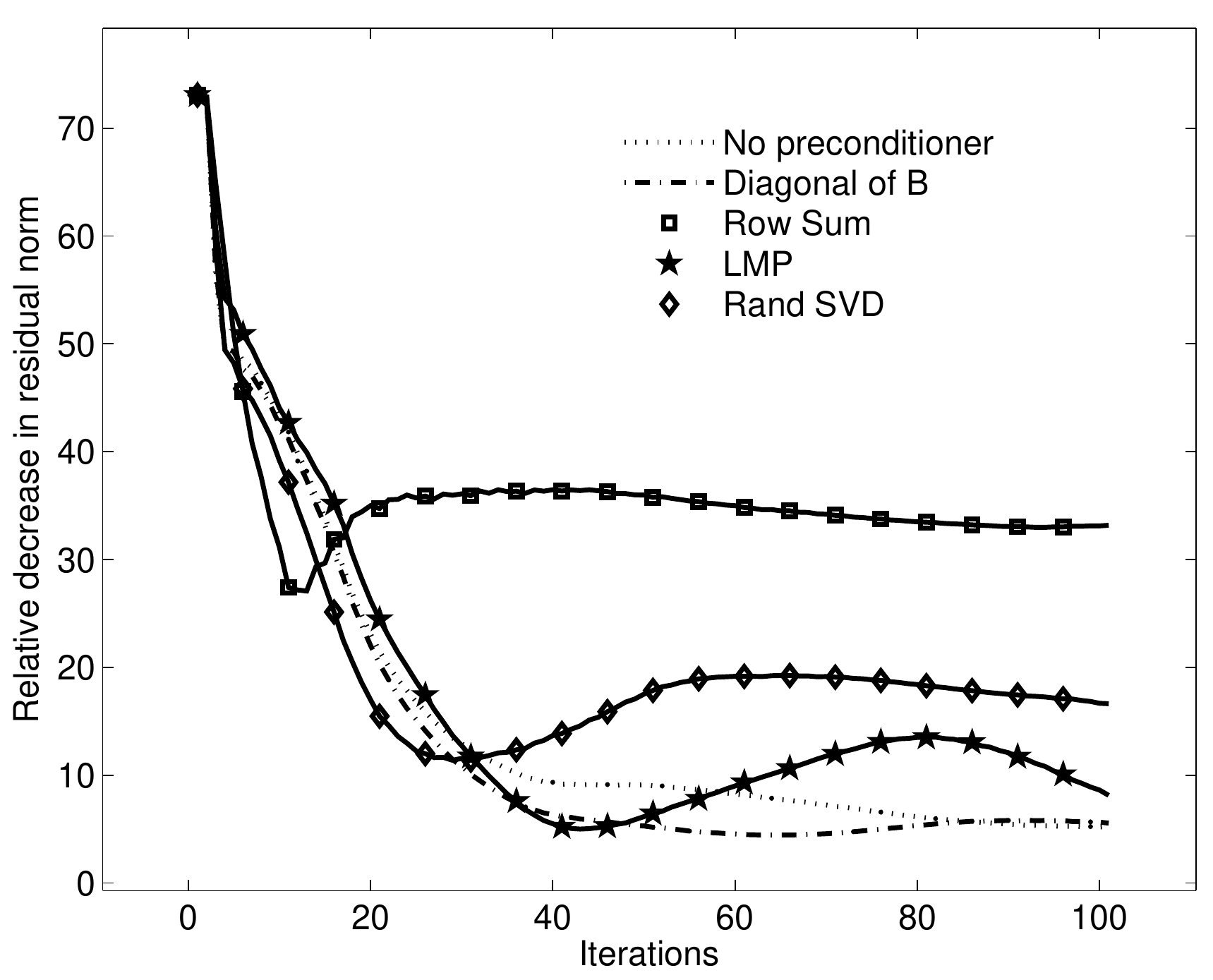}
\label{fig:wrfprec}
}
\caption{ Convergence of preconditioned iterative solvers for the ({\sc wrf}) supersensitivity system \eqref{eqn:linsys}.}
\end{figure}

To solve the linear system associated with WRF we use the GMRES algorithm from the PETSc software library, since this algorithm can handle nonsymmetric and indefinite matrices.  We select a subset of the preconditioners used with the {\sc swe} model. The first preconditioner (and the easiest to obtain) is the diagonal of the covariance matrix $\B_0$. The second preconditioner is the sum of elements in each row. The third preconditioner is a limited memory quasi-Newton approximation that uses information gathered throughout the data assimilation process. As shown in \cite{Tshimanga_2008}, the descent directions generated by the minimizer can be used to build the limited memory preconditioner through the L-BFGS formula. The fourth and last preconditioner used is the randomized SVD with $100$ random vectors, computed in parallel at the equivalent total cost of just two model runs. The decrease in the norm of residual is presented in Figure \ref{fig:wrfprec} and in Table \ref{tab:wrfprec}.

\begin{table}[ht]
\centering
\begin{tabular}{|c||c|c|c|c|c|}
\hline
Preconditioner & Relative decrease \\
               &  in residual norm \\
\hline \hline
None           & $7.2$e-2  \\
\hline
Background     & $7.6$e-2  \\
\hline
Row sum        & $4.5$e-1  \\
\hline
LMP            & $1.1$e-1  \\
\hline
Randomized SVD & $2.2$e-1  \\
\hline
\end{tabular}
\caption{ Solution residual norm after 100 preconditioned iterations of GMRES for the ({\sc wrf}) supersensitivity system \eqref{eqn:linsys}. The scaling is done with respect to the initial guess residual norms. }
\label{tab:wrfprec}
\end{table}

As we can see from these results, the convergence of GMRES did not improve considerably through preconditioning. Moreover, while the unpreconditioned solver reduces the error of the residual monotonically, the preconditioned ones do not. The row sum preconditioner performs better than all the others in the first $15$ iterations, then starts departing from the solution. A similar behavior can be observed for the preconditioner obtained from randomized SVD, which performs best between the $15$-th and $30$-th iterations. The diagonal of $\B_0$ preconditioner is the best for the next $50$ iterations, except for a small interval where the LMP is slightly better. After $100$ iterations the unpreconditioned residual is the smallest. In conclusion, it is really difficult to pinpoint one particular preconditioner as performing best for our WRF model. 
The fact that each solver leads to a residual that first decreases, then starts to increase requires further investigation. We think that this behavior is due to the large approximation errors made in computing first and second-order information. We are working with a 4D-Var reanalysis that is not optimal, and with adjoint models that are incomplete. Moreover, we employ Gauss-Newton approximation of the 4D-Var Hessian, and the ignored higher order terms may be non-negligible at the suboptimal solution. Other errors are associated with the way WRF deals with boundary conditions. Our methodology is affected by all these factors and the problem cannot be solved to a high degree of accuracy without improving the quality of each of these elements. 


\section{Visual Analysis of Sensitivity Results}\label{sec:vis}

In this section we illustrate the sensitivity analysis results. Consider the {\sc swe} data assimilation test case described in Section \ref{sec:das}, except two of the observations are faulty. The sensitivity analysis results should reflect this inconsistency in observations. 

\begin{figure}[ht]
\centering

\subfigure[Observations for $h$]{
\includegraphics[height=0.45\textwidth,width=0.65\textwidth]{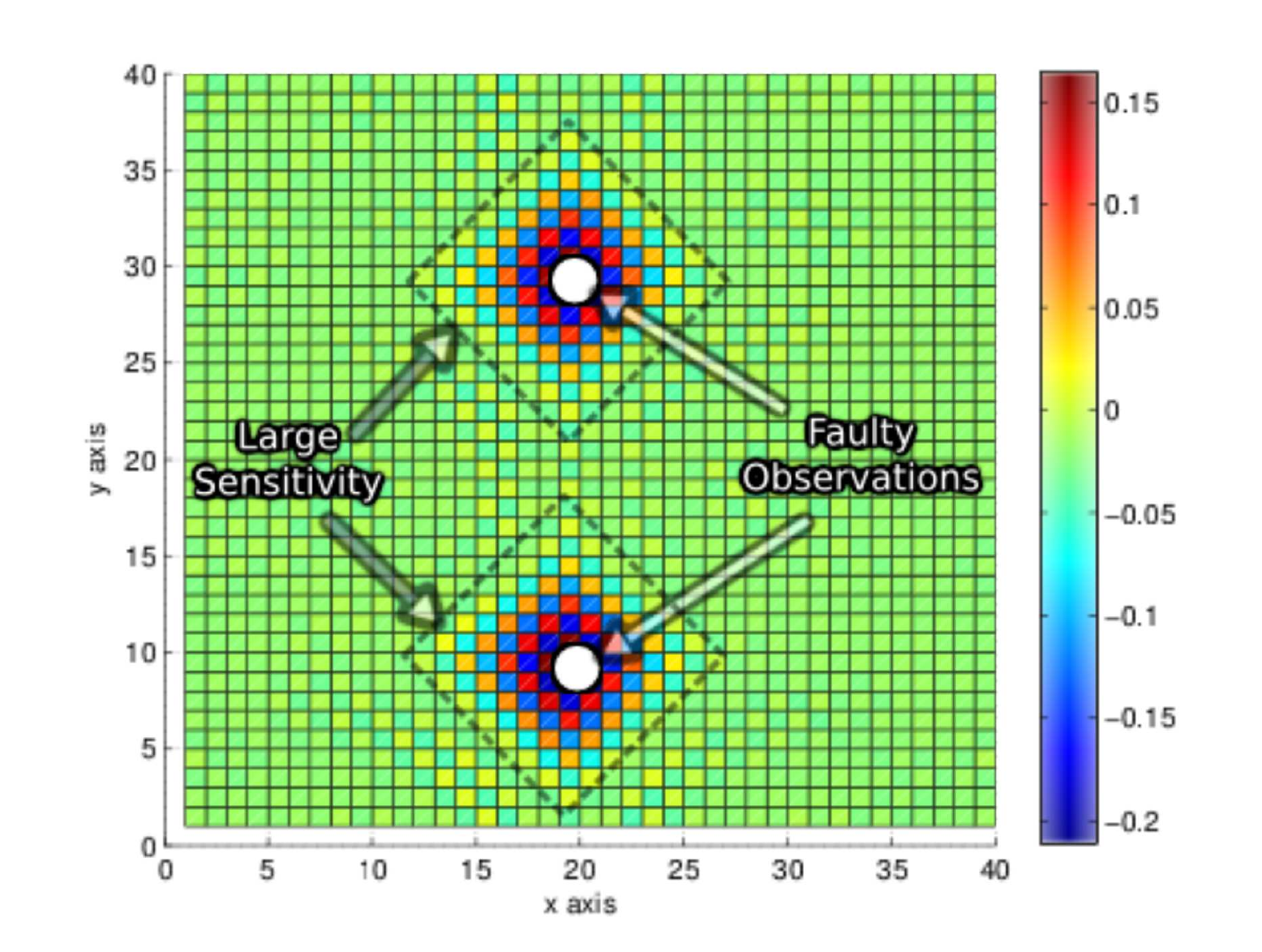}
\label{fig:sns9h}
}
\subfigure[Observations for $u$]{
\includegraphics[height=0.45\textwidth,width=0.65\textwidth]{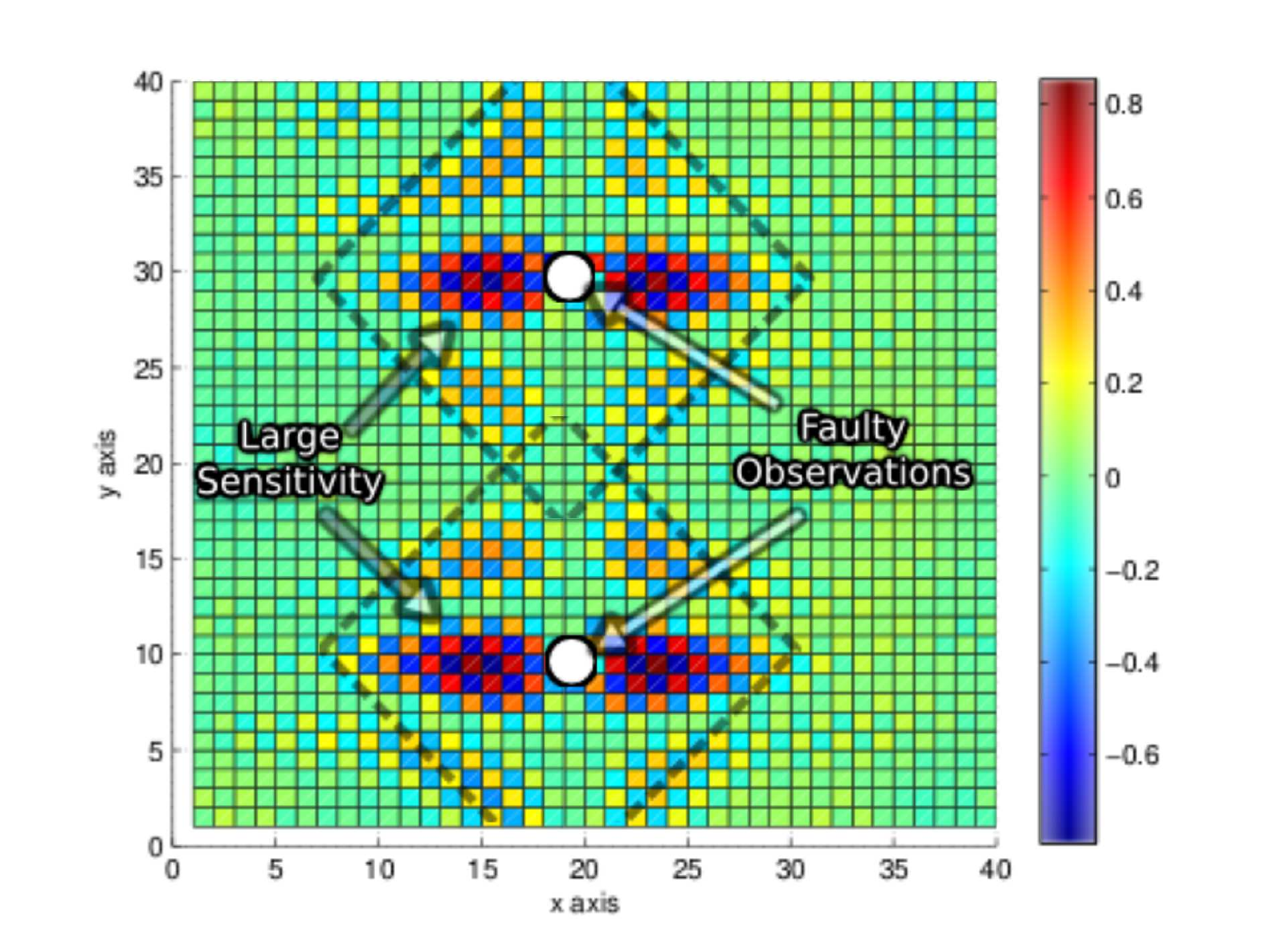}
\label{fig:sns9u}
}
\subfigure[Observations for $v$]{
\includegraphics[height=0.45\textwidth,width=0.65\textwidth]{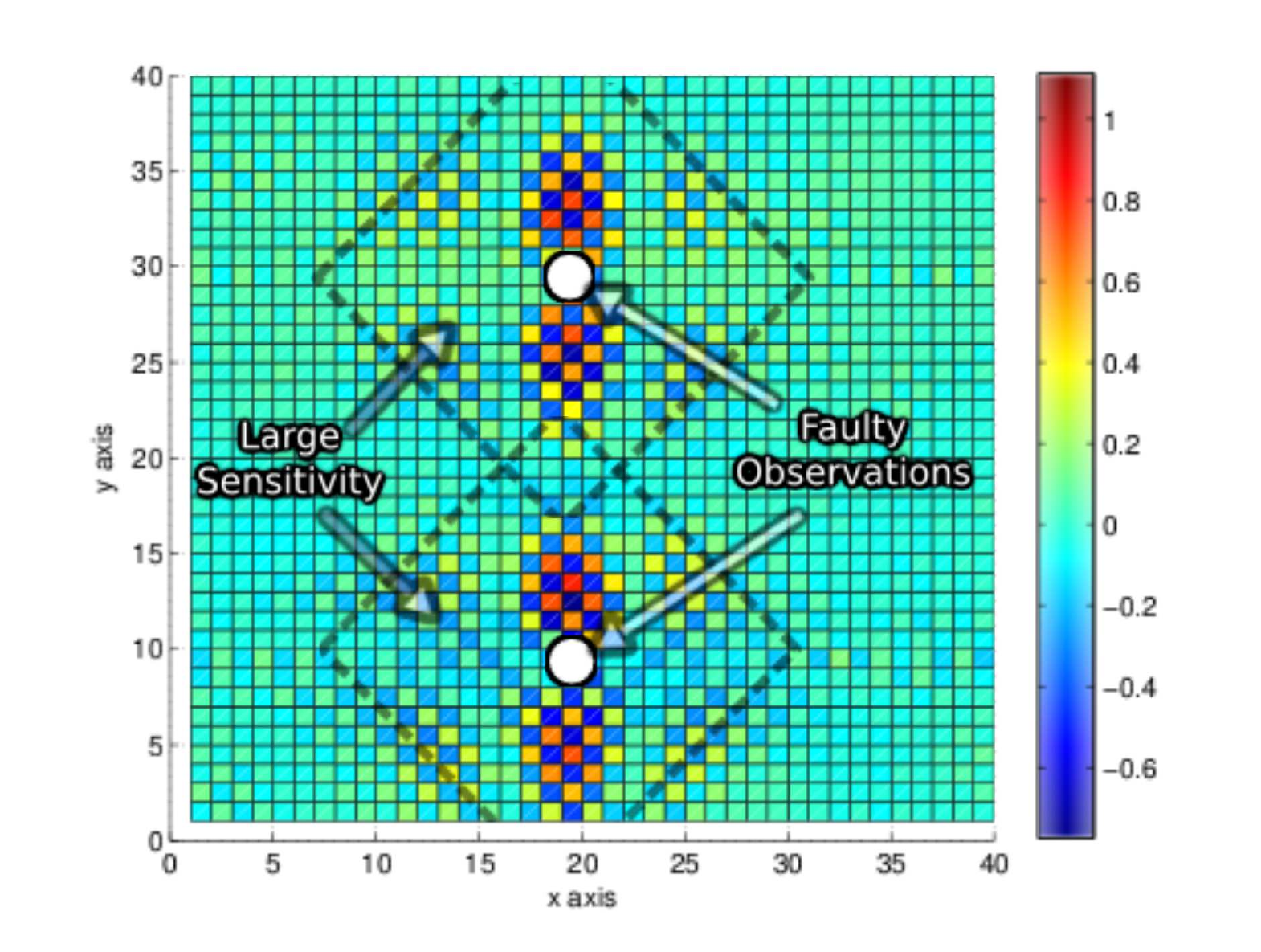}
\label{fig:sns9v}
}
\caption{\label{fig:sns9}Fields of forecast sensitivities to observations, represented on the computational grid.}
\end{figure}

Our approach is to modify the value of observations corresponding to $h$, $u$, $v$ at two locations, before starting the assimilation process. This is done only for the final time of the assimilation window. The modified observations are located on the North-South median line, at coordinates $10$x$20$ and $30$x$20$ on the $40$x$40$ grid, as shown in Figure \ref{fig:sns9}. The two locations were chosen to be isolated from each other so that the associated sensitivities will have a smaller chance of totally overlapping. Due to the symmetry of the locations, it is expected the results will be easier to study intuitively. 

The fields of supersensitivities corresponding to $h$, $u$ and $v$ are plotted in Figures \ref{fig:sns9h}, \ref{fig:sns9u}, \ref{fig:sns9v}. The sensitivities have nonzero values and a pulse-like structure centered at the grid points containing the faulty observations. This indicates that the forecast error is most sensitive to the data assimilation parameters defined on these areas, such as the faulty observations. Although we modified the value of observations at two individual sites, the sensitivities are shaped as a pulse because the correlation between model variables spreads the errors spatially. When passing the supersensitivity through the {\sc TLM} model to obtain the sensitivity to parameters defined at future times, the shape and location of the sensitivity is preserved (not shown here). This confirms the theory of 4D-Var that the information (or errors) in the observations are also spread in time. 

\section{Conclusions}\label{sec:end}

In data assimilation the sensitivity of a forecast aspect to observations
provides a quantitative metric of the impact each data point has on reducing forecast uncertainty. 
This metric can be used in hindsight to prune redundant data, to identify faulty measurements, and to improve the parameters of the data assimilation system. The metric
can also be used in foresight to  adaptively configure, and deploy, sensor networks for future measurements. 

This work provides a systematic study of computational strategies to obtain sensitivities to observations in the context of 4D-Var data assimilation. 
Solution efficiency is
of paramount importance since the models of interest in practice are large scale, and
the computational cost of sensitivities is considerable;  moreover, in an operational setting, the sensitivities have to be solved in 
faster-than-real-time (e.g., for dynamically deploying new sensors). 

The cost of computing sensitivities to observations is dominated by the solution of a large-scale linear system, whose matrix is the Hessian of the 4D-Var cost function. In practice, this matrix is available only in operator form (i.e., matrix-vector products obtained via second order adjoint models). 

The main contributions of this paper are to formulate the computational challenges associated with sensitivities to observations, and to present solutions to address them. We consider a set of matrix-free linear solvers, build specific preconditioners, and compare their performance on two numerical models. For the {\sc swe} test, the results are very promising: certain preconditioners as well as the multigrid approach lead to significant efficiency improvements in the solution of the linear system. The results for the WRF test are less clear cut: preconditioning brings only a modest improvement, and we attribute this to the limited accuracy with which derivatives are computed by the (currently incomplete) WRF adjoint model.
Future work with WRF should focus both on finding better preconditioners, and on developing a more accurate adjoint model. 
%


%
%

%

\subsubsection*{Acknowledgments.} This work was supported by National Science Foundation through the awards NSF DMS-0915047, NSF CCF-0635194, NSF CCF-0916493 and NSF OCI-0904397; and by AFOSR through the award FA9550--12--1--0293--DEF.

\newpage

\bibliographystyle{elsarticle-num}
\bibliography{obsimp_journal}

\end{document}